\PassOptionsToPackage{unicode}{hyperref}
\PassOptionsToPackage{hyphens}{url}
\documentclass[
  conference]{IEEEtran}
\usepackage{xcolor}
\usepackage{amsmath,amssymb}
\usepackage{iftex}
\ifPDFTeX
  \usepackage[T1]{fontenc}
  \usepackage[utf8]{inputenc}
  \usepackage{textcomp} 
\else 
  \usepackage{unicode-math} 
  \defaultfontfeatures{Scale=MatchLowercase}
  \defaultfontfeatures[\rmfamily]{Ligatures=TeX,Scale=1}
\fi
\usepackage{lmodern}
\ifPDFTeX\else
\fi
\IfFileExists{upquote.sty}{\usepackage{upquote}}{}
\IfFileExists{microtype.sty}{
  \usepackage[]{microtype}
  \UseMicrotypeSet[protrusion]{basicmath} 
}{}
\makeatletter
\@ifundefined{KOMAClassName}{
  \IfFileExists{parskip.sty}{%
    \usepackage{parskip}
  }{
    \setlength{\parindent}{0pt}
    \setlength{\parskip}{6pt plus 2pt minus 1pt}}
}{
  \KOMAoptions{parskip=half}}
\makeatother
\usepackage{graphicx}
\makeatletter
\newsavebox\pandoc@box
\newcommand*\pandocbounded[1]{
  \sbox\pandoc@box{#1}%
  \Gscale@div\@tempa{\textheight}{\dimexpr\ht\pandoc@box+\dp\pandoc@box\relax}%
  \Gscale@div\@tempb{\linewidth}{\wd\pandoc@box}%
  \ifdim\@tempb\p@<\@tempa\p@\let\@tempa\@tempb\fi
  \ifdim\@tempa\p@<\p@\scalebox{\@tempa}{\usebox\pandoc@box}%
  \else\usebox{\pandoc@box}%
  \fi%
}
\def\fps@figure{htbp}
\makeatother
\NewDocumentCommand\citeproctext{}{}

\makeatletter
 \let\@cite@ofmt\@firstofone
 \def\@biblabel#1{}
 \def\@cite#1#2{{#1\if@tempswa , #2\fi}}
\makeatother
\newlength{\cslhangindent}
\newlength{\csllabelwidth}
\newenvironment{CSLReferences}[2] 
 {\begin{list}{}{%
  \setlength{\itemindent}{0pt}
  \setlength{\leftmargin}{0pt}
  \setlength{\parsep}{0pt}
  \ifodd #1
   \setlength{\leftmargin}{\cslhangindent}
   \setlength{\itemindent}{-1\cslhangindent}
  \fi
  \setlength{\itemsep}{#2\baselineskip}}}
 {\end{list}}
\usepackage{calc}

\providecommand{\tightlist}{%
  \setlength{\itemsep}{0pt}\setlength{\parskip}{0pt}}
\usepackage{booktabs}
\usepackage{array}
\usepackage{hyperref}
\usepackage{bookmark}
\IfFileExists{xurl.sty}{\usepackage{xurl}}{} 
\makeatletter
\@ifundefined{xmpquote}{}{}
\makeatother
\hypersetup{
  pdftitle={A Synthetic Benchmark Dataset with Endogenous Marketing Spend for Validating Marketing Mix Models},
  pdfauthor={Niklas Heusch (niklas.heusch@zalando.de)},
  hidelinks,
  pdfcreator={LaTeX via pandoc}}

\title{A Synthetic Benchmark Dataset with Endogenous Marketing Spend for
Validating Marketing Mix Models}
\author{Niklas Heusch (niklas.heusch@zalando.de)}
\date{Draft --- September 2026}

\begin{document}
\maketitle
\begin{abstract}
Marketing Mix Models (MMMs) estimate the incremental sales effect of
advertising from observational time series, yet they are rarely
validated against ground truth, because ground truth is unobservable in
real data. Synthetic data closes that gap in principle, but existing
generators produce marketing spend exogenously --- omitting the central
difficulty of the estimation problem, since real budgets are planned
around promotional calendars, seasons, and recent performance. This
paper presents a parameterized generator, and a fixed reference
instance, of a synthetic weekly retail dataset (156 weeks, three media
channels) in which spend arises from four documented coordination
mechanisms --- quarterly budget feedback, anticipatory spending ahead of
a promotional calendar, scheduled TV bursts, and algorithmic performance
chasing --- on a demand baseline with seasonal, quality, price, and
unobserved sentiment components. Spend translates into incremental sales
through two transformations, carryover and diminishing returns,
instantiated here as geometric adstock and logistic saturation with
known parameters; the true causal decomposition of every week's sales is
recorded alongside the error-contaminated variables a practitioner would
observe. Every mechanism is a parameter that can be varied or switched
off, and a companion procedure simulates go-dark geo-experiments with
exact treatment effects. The seeded generator and reference instance are
publicly released with notebooks that reproduce every number in this
paper.
\end{abstract}

\textbf{Companion materials:}
\href{https://github.com/niklas-heusch/mmm-materials}{Code and
notebooks} ·
\href{https://niklas-heusch.github.io/posts/mmm1-synthetic-data/mmm1-synthetic-data.html}{Illustrated
walkthrough}

\section{Introduction}\label{introduction}

Marketing Mix Modeling has returned to prominence as privacy regulation
and platform policy have degraded user-level attribution. Open-source
frameworks --- Meta's Robyn (Runge et al. 2023), Google's Meridian
(Zhang et al. 2024), and pymc-marketing (Orduz 2024) --- have lowered
the cost of fitting an MMM to the point where a single analyst can
produce channel-level estimates of return on advertising spend (ROAS) in
an afternoon. The estimates inform budget decisions that, for a
mid-sized retailer, run to tens of millions of euros per year.

The fundamental problem is that these estimates are difficult to
validate. The quantity an MMM claims to measure --- incremental sales
caused by each channel --- is never observed in real data. In practice,
validation reduces to predictive fit, face validity, or comparison with
other MMM estimates, none of which establishes that the causal
decomposition is correct. A model can fit the sales series well while
attributing seasonal demand to the media channels that happen to spend
into it.

Synthetic data offers a direct solution: generate data from a known
causal structure, fit the estimator on the variables a practitioner
would observe, and check whether it recovers the known truth. Several
such generators exist. However, they share a limitation that undermines
their value as benchmarks: they generate marketing spend exogenously ---
as random draws, user-specified schedules, or simple seasonal patterns.
This removes precisely the feature that makes MMM estimation difficult
in practice. Budget-setting in real firms is deliberate and
forward-looking: marketing teams schedule campaigns around promotional
calendars and seasons known months in advance, bidding algorithms raise
spend when recent performance is strong, and finance reallocates budgets
each quarter in response to sales. Each of these behaviors ties spend to
the determinants of sales through channels other than advertising's own
effect. A benchmark without them tests an estimator on a version of the
problem from which the hard part has been removed.

This paper describes a synthetic dataset constructed to close that gap
--- a dataset that is realistic in a specific, documented sense: spend
is generated by the coordination mechanisms firms actually use. The
contributions are:

\begin{enumerate}
\def\labelenumi{\arabic{enumi}.}
\tightlist
\item
  \textbf{A data-generating process (DGP) with explicit, parameterized
  coordination.} Marketing spend responds to (i) quarterly budget
  reviews tied to recent business performance, (ii) an anticipated
  promotional calendar, (iii) a planned TV burst schedule, and (iv)
  lagged sales performance through a stylized algorithmic bidding rule.
  Each mechanism is a separate, documented function in the generator and
  can be varied or switched off, supporting ablation studies.
\item
  \textbf{A dual observed/ground-truth structure.} The dataset records
  both the true causal decomposition of sales (baseline, promotional
  uplift, per-channel media effect, week by week) and the
  error-contaminated variables a practitioner would actually observe,
  including a promotion indicator with realistic misclassification and a
  price series with autocorrelated measurement error. One demand
  component --- market sentiment --- is deliberately excluded from the
  observed set.
\item
  \textbf{A geo-experiment extension.} A companion procedure simulates
  ``go-dark'' geo-experiments consistent with the same DGP, with the
  exact treatment effect known in every week, including the carryover
  after each test ends.
\end{enumerate}

Both the parameterized generator and a fixed, seeded reference instance
are released, together with notebooks that reproduce the instance
bit-for-bit and every figure and number in this paper. All data is
synthetic: parameter values, magnitudes, and business mechanisms are
fictional, chosen to lie in the range of published industry benchmarks,
and no quantity derives from any real company's data.

Section 2 relates the dataset to existing simulators and benchmarks.
Section 3 specifies the DGP. Section 4 describes the reference instance.
Section 5 presents the geo-experiment extension. Section 6 discusses
intended use and limitations, and Section 7 states availability.

\section{Related work}\label{related-work}

Two open-source simulators are directly comparable. Meta's siMMMulator
(Nguyen 2022) generates weekly MMM input data from first principles ---
baseline sales, spend, impressions, conversions --- with known channel
returns, explicitly for validating and comparing MMMs. Spend, however,
is specified by the user or drawn from simple distributions; the package
does not model feedback from sales to spend. Google's Aggregate
Marketing System Simulator (Vaver and Zhang 2017) is considerably
richer: an agent-based model in which consumers move through mindset
states in response to marketing and seasonality. AMSS can represent
sophisticated targeting behavior, but its complexity makes the mapping
from parameters to induced confounding opaque, and the R package has
seen limited recent development. The present dataset occupies a middle
position: simple enough that every source of confounding is a named term
in a closed-form equation, rich enough that the main coordination
mechanisms observed in retail marketing operations are present.

Framework documentation also relies on simulated examples --- Robyn
ships a simulated weekly dataset, and pymc-marketing's tutorials
generate their own --- but these serve as illustrations rather than
benchmarks, and again treat spend as exogenous. Zhang et al. (2024) use
simulation to study MMM calibration with Bayesian priors, generating
confounding through omitted variables rather than through an explicit
spend-setting process.

The validation problem itself is well documented. Lewis and Rao (2015)
show that the sales impact of advertising is small relative to the
variance of sales, making observational estimates fragile. Blake et al.
(2015) find, using large-scale experiments at eBay, that observational
estimates of paid-search effectiveness were severely inflated. Gordon et
al. (2019) compare observational methods against randomized experiments
across Facebook campaigns and conclude that observational methods
frequently fail to recover experimental lift. These studies motivate
treating recovery of a known truth --- rather than predictive fit --- as
the standard of validation, which is the purpose synthetic data serves.
On the experimental side, Vaver and Koehler (2011) introduce the
geo-experiment framework that Section 5 simulates.

\section{Data-generating process}\label{data-generating-process}

\subsection{Overview and notation}\label{overview-and-notation}

\begin{figure}
\centering
\pandocbounded{\includegraphics[keepaspectratio,alt={The data-generating process at a glance. Boxes are the DGP's components; the dashed border marks market sentiment, which has no observed counterpart. Black arrows are the demand and media-response structure; yellow arrows are the two most consequential coordination mechanisms --- spend rises ahead of calendar events (anticipation) and follows lagged pre-media sales (quarterly budgets and the bidding algorithm). Planned spend additionally tilts mildly with the season and follows TV's burst calendar; all four mechanisms are specified in Section 3.4.}]{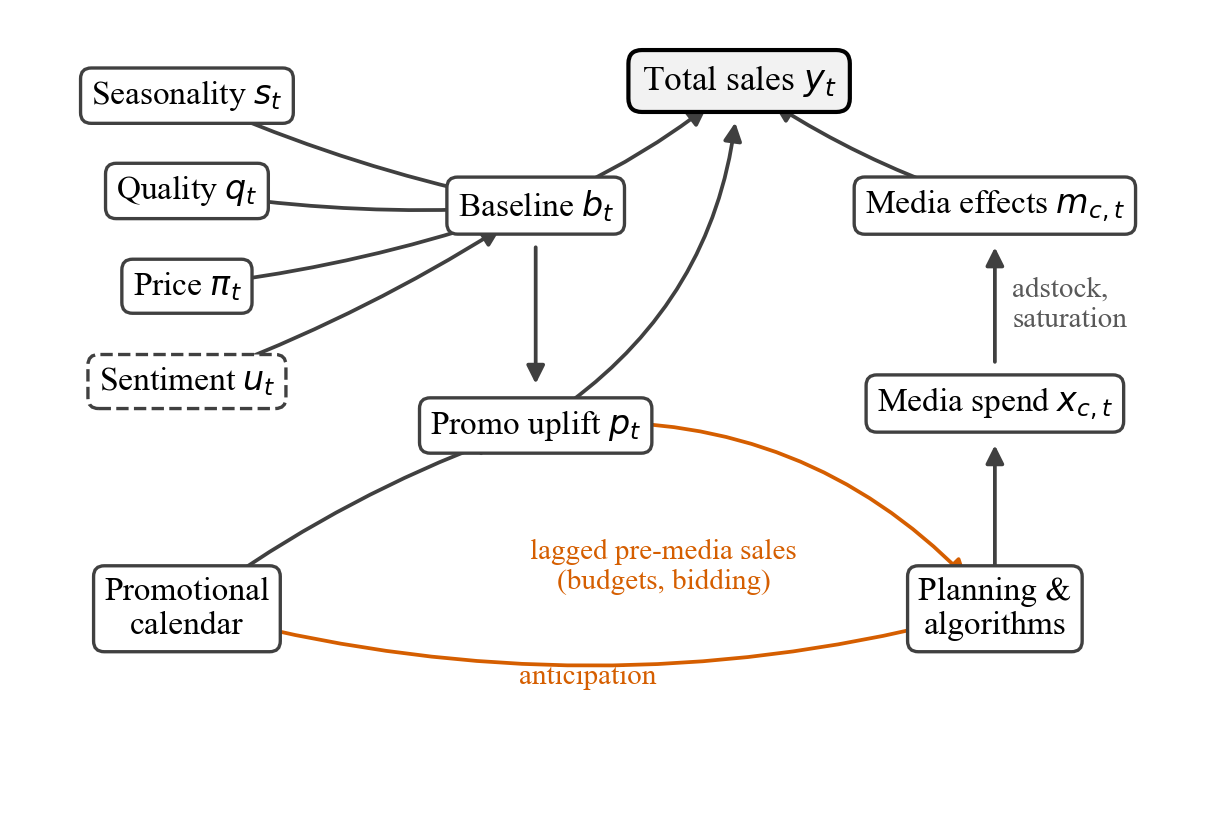}}
\caption{The data-generating process at a glance. Boxes are the DGP's
components; the dashed border marks market sentiment, which has no
observed counterpart. Black arrows are the demand and media-response
structure; yellow arrows are the two most consequential coordination
mechanisms --- spend rises ahead of calendar events (anticipation) and
follows lagged pre-media sales (quarterly budgets and the bidding
algorithm). Planned spend additionally tilts mildly with the season and
follows TV's burst calendar; all four mechanisms are specified in
Section 3.4.}\label{fig:dgp}
\end{figure}

The DGP produces weekly observations \(t = 1, \dots, T\) with
\(T = 156\) (three years). All monetary quantities are stored in
thousands of euros; prose and figures quote weekly sales in millions of
euros and media spend in thousands, whichever yields compact numbers.
Total sales decompose additively as

\[y_t = b_t + p_t + \sum_{c} m_{c,t}, \tag{1}\]

where \(b_t\) is baseline sales absent promotions and marketing, \(p_t\)
is promotional uplift, and \(m_{c,t}\) is the incremental effect of
media channel \(c \in \{\text{PLA}, \text{Meta}, \text{TV}\}\) (PLA
denotes product listing ads, i.e., paid shopping). Equation (1) defines
the ground truth, and every term on its right-hand side is a column of
the dataset. Generation proceeds in five steps --- baseline, promotions,
spend, media effects, and measurement --- summarized in
Figure~\ref{fig:dgp}. The yellow arrows are the point of the exercise:
they are the paths through which spend responds to the determinants of
sales.

\subsection{Baseline sales}\label{baseline-sales}

Baseline sales combine four components:

\[b_t = B \,(1 + s_t)\,(1 + q_t)\,\pi_t^{-0.9} + B\, u_t + \varepsilon_t, \tag{2}\]

with base level \(B = 2{,}000\) (i.e., €2 million per week), i.i.d.
noise \(\varepsilon_t \sim \mathcal{N}(0, (0.02B)^2)\), and a floor at
\(0.5B\) as a guardrail. The components are:

\begin{itemize}
\tightlist
\item
  \textbf{Seasonality}
  \(s_t = 0.15\sin(2\pi t/52) + 0.08\sin(4\pi t/52) + 0.05\sin(6\pi t/52)\):
  three harmonics with periods of 52, 26, and 17.3 weeks and amplitudes
  of 15, 8, and 5 percent.
\item
  \textbf{Product quality} \(q_t\): a persistent AR(1) process,
  \(q_t = 0.95\, q_{t-1} + \nu_t\),
  \(\nu_t \sim \mathcal{N}(0, 0.02^2)\), representing gradual assortment
  changes.
\item
  \textbf{Competitive price position} \(\pi_t\): an AR(1) process
  reverting to 1, \(\pi_t = 0.9\,\pi_{t-1} + 0.1 + \omega_t\),
  \(\omega_t \sim \mathcal{N}(0, 0.03^2)\), entering through a base
  price elasticity of \(-0.9\).
\item
  \textbf{Market sentiment} \(u_t\): an AR(1) process,
  \(u_t = 0.7\, u_{t-1} + \xi_t\),
  \(\xi_t \sim \mathcal{N}(0, 0.03^2)\), initialized at
  \(\mathcal{N}(0, 0.05^2)\). Sentiment is deliberately excluded from
  the observed variables (see the measurement layer below); it is the
  dataset's unobserved demand shifter.
\end{itemize}

The mixed structure of Equation (2) --- seasonality and quality scale
the business multiplicatively, sentiment enters additively --- is
deliberate: real demand components rarely combine in the neat additive
way estimation models assume, so an estimator that controls for them
linearly faces a mild functional-form mismatch even when the components
are observed.

\subsection{Promotional calendar and price
response}\label{promotional-calendar-and-price-response}

Promotions follow a fixed annual calendar of five events
(Table~\ref{tbl:calendar}), repeated in each of the three years --- the
predictability that allows marketing teams to plan around them.

\begin{table}[!htbp]
\centering
\footnotesize
\caption{Promotional calendar.}\label{tbl:calendar}
\begin{tabular}{@{}lccc@{}}
\toprule
Event & Weeks of year & Discount & Elasticity multiplier \\
\midrule
Winter clearance & 2–4 & 40\% & 1.5 \\
Spring mid-season & 16 & 30\% & 1.3 \\
Summer clearance & 28–29 & 40\% & 1.5 \\
Autumn mid-season & 40 & 30\% & 1.3 \\
Cyber week & 46–47 & 40\% & 1.8 \\
\bottomrule
\end{tabular}
\par\vspace{0.8em}
\begin{minipage}{\columnwidth}
\footnotesize
\textit{Note:} Repeated in each of the three years. The elasticity multiplier scales the base price elasticity of $-0.9$ during the event, reflecting that deal-seeking customers respond more strongly to a given discount.
\end{minipage}
\end{table}

The promotional sales multiplier and uplift are

\[\mu_t = \big(\max(1 - d_t,\, 0.5)\big)^{\,e_t}, \qquad p_t = b_t\,(\mu_t - 1), \tag{3}\]

where \(d_t\) is the discount and \(e_t\) the occasion-scaled
elasticity; \(\mu_t\) is clipped to \([0.8, 3.0]\), guardrails that
never bind in the reference instance. Because uplift is proportional to
\(b_t\), promotional effects interact with the baseline --- a second
deliberate departure from the additive-in-controls form.

\subsection{Marketing spend: four coordination
mechanisms}\label{marketing-spend-four-coordination-mechanisms}

Planned spend is generated on the log scale. For channel \(c\) in week
\(t\),

\[\log x^{\text{plan}}_{c,t} = \log \bar{x}_c + \log Q_t + 0.15\, s_t + \gamma_c A_t + \tau_{c,t} + \eta_{c,t}, \tag{4}\]

with planning noise \(\eta_{c,t} \sim \mathcal{N}(0, 0.1^2)\) and base
weekly levels \(\bar{x}_c\) of €60K (PLA), €40K (Meta), and €10K (TV).
Realized spend equals planned spend, except for PLA, where an
algorithmic multiplier applies on top:
\(x_{\text{pla},t} = x^{\text{plan}}_{\text{pla},t} \cdot \rho_t\). The
terms implement the coordination mechanisms:

\textbf{Quarterly budget feedback} \(Q_t\). From the first review with
two full quarters of history onward, budgets adjust every 13 weeks to
the ratio of the previous quarter's pre-media sales to the quarter
before:

\[Q_t = \operatorname{clip}\!\big(1 + 0.3\,(r^{Q}_t - 1),\; 0.8,\; 1.4\big), \tag{5}\]

held constant within the quarter. In the reference instance the realized
multipliers range from 0.90 to 1.18 (Figure~\ref{fig:endogeneity}, panel
C). This creates low-frequency correlation between past demand
conditions and current spend across all channels.

\textbf{Anticipatory promotional spending} \(A_t\). \(A_t\) counts
promotional weeks among \(t+1, \dots, t+3\); the loading \(\gamma_c\) is
0.4 for PLA and Meta and 0 for TV. Spend therefore rises \emph{before}
demand events, not in response to them --- a forward-looking form of
coordination that lag-based controls do not remove. In the reference
instance, mean Meta spend is €73.3K in weeks with at least one promotion
among the next three, versus €40.9K in other weeks
(Figure~\ref{fig:endogeneity}, panel A).

\textbf{Scheduled TV bursts} \(\tau_{c,t}\). TV follows a flighting
pattern: a five-week pre-Christmas burst (spend multiplied by 15) and a
three-week early-summer burst (multiplied by 8) each year, reflecting
long-lead-time brand campaigns concentrated in a few weeks (visible in
Figure~\ref{fig:overview}, panel B).

\textbf{Algorithmic performance chasing} \(\rho_t\) (PLA only). A
stylized bidding rule compares last week's sales to a ten-week rolling
baseline and scales spend by

\[\rho_t = \operatorname{clip}\!\big(1 + 0.8\,(r_t - 1),\; 0.5,\; 3.0\big), \qquad r_t = \frac{y^{\ast}_{t-1}}{\tfrac{1}{10}\sum_{j=2}^{11} y^{\ast}_{t-j}}, \tag{6}\]

where \(y^{\ast}\) denotes sales \emph{before} media effects (baseline
plus promotional uplift). The rule captures the defining feature of
automated bidding --- raising spend when recent performance is strong
--- and induces a correlation of 0.46 between PLA spend and the lagged
performance ratio in the reference instance
(Figure~\ref{fig:endogeneity}, panel B). Using pre-media sales as the
signal is a documented simplification: a production algorithm observes
total sales, including its own effect, which would create a feedback
loop between Equations (4) and (8). We accept the simplification to keep
the causal graph acyclic and the ground truth exactly computable; it
understates, rather than overstates, the coordination between PLA spend
and demand.

\subsection{Media effects}\label{media-effects}

Marketing spend translates into incremental sales through two
transformations. The first, \emph{carryover}, spreads a week's
advertising effect over subsequent weeks; the second, \emph{diminishing
returns}, makes each additional euro within a week buy less than the
last. We instantiate carryover as geometric adstock with a six-week
window and weights normalized to sum to one,

\[\tilde{x}_{c,t} = \frac{\sum_{\ell=0}^{5} \alpha_c^{\ell}\, x_{c,t-\ell}}{\sum_{\ell=0}^{5} \alpha_c^{\ell}}, \tag{7}\]

so that total effect is redistributed over time rather than amplified,
and diminishing returns as logistic saturation,

\[g(\tilde{x}; \lambda) = \frac{1 - e^{-\lambda \tilde{x}}}{1 + e^{-\lambda \tilde{x}}}, \qquad m_{c,t} = \beta_c\, g(\tilde{x}_{c,t}; \lambda_c). \tag{8}\]

\begin{figure}
\centering
\pandocbounded{\includegraphics[keepaspectratio,alt={True steady-state response curves: incremental weekly sales against adstocked weekly spend, \textbackslash beta\_c\textbackslash, g(\textbackslash tilde\{x\}; \textbackslash lambda\_c). Because the adstock weights sum to one, a constant weekly spend of x has adstocked spend x, so the curve also reads as the steady-state return to a constant budget. Dots mark each channel's mean adstocked spend --- its typical operating point --- annotated with the average return there; the dotted line is break-even.}]{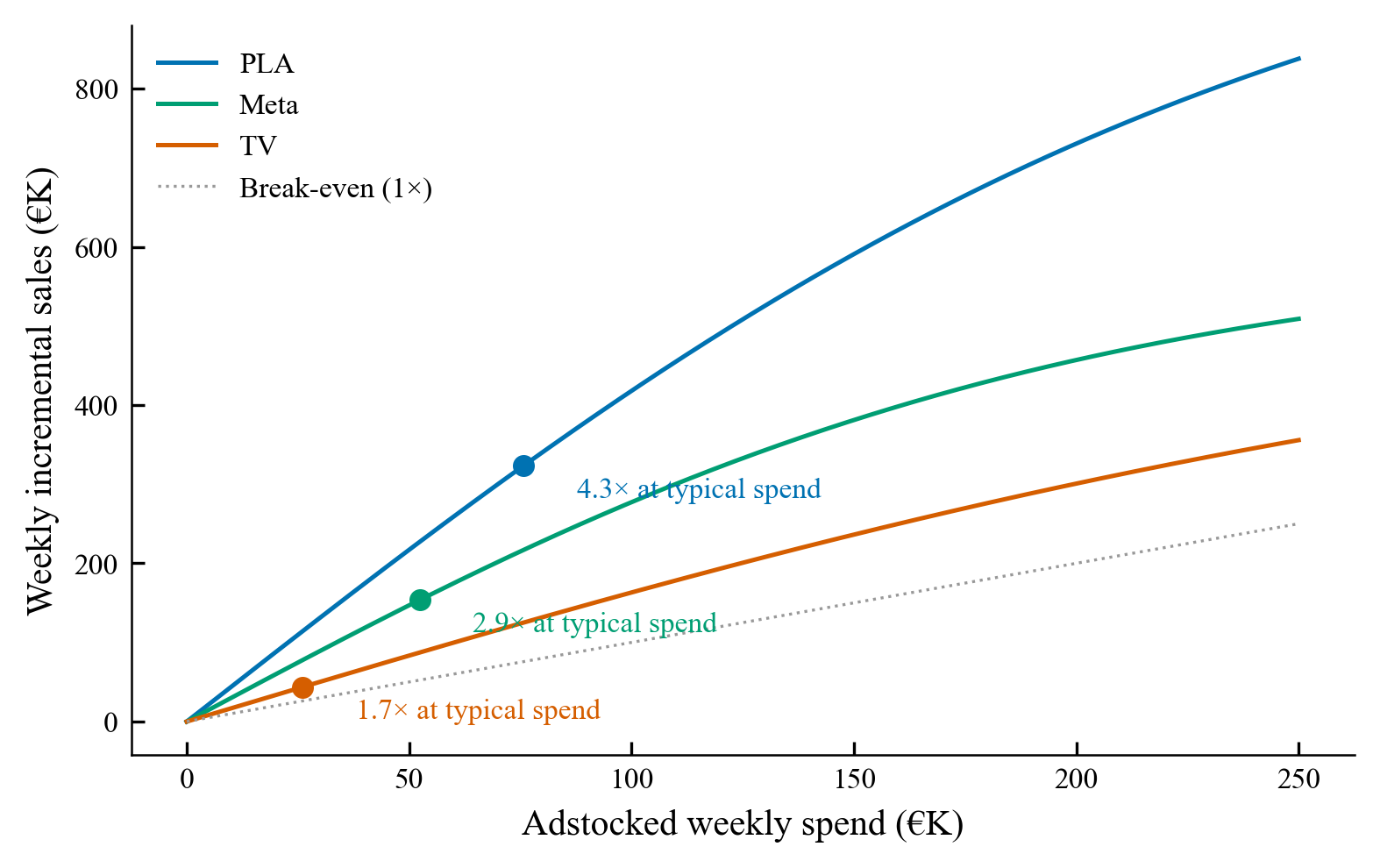}}
\caption{True steady-state response curves: incremental weekly sales
against adstocked weekly spend, \(\beta_c\, g(\tilde{x}; \lambda_c)\).
Because the adstock weights sum to one, a constant weekly spend of \(x\)
has adstocked spend \(x\), so the curve also reads as the steady-state
return to a constant budget. Dots mark each channel's mean adstocked
spend --- its typical operating point --- annotated with the average
return there; the dotted line is break-even.}\label{fig:response-curves}
\end{figure}

Both are common choices among several in use --- Weibull adstock and
Hill saturation are equally standard --- and each is a single function
in the generator, straightforward to replace. Channel parameters and the
implied returns are given in Table~\ref{tbl:channels}; ROAS is defined
as \(\sum_t m_{c,t} / \sum_t x_{c,t}\), the average return over the
sample. Figure~\ref{fig:response-curves} shows the implied response
curves with each channel's operating point: PLA operates well into the
concave region of its saturation curve, while TV --- outside its bursts
--- operates near the origin, where returns are close to linear.

\begin{table}[!htbp]
\centering
\footnotesize
\caption{True channel parameters (reference instance).}\label{tbl:channels}
\begin{tabular}{@{}lccc@{}}
\toprule
 & PLA & Meta & TV \\
\midrule
Adstock $\alpha_c$ & 0.2 & 0.4 & 0.7 \\
Saturation $\lambda_c$ & 0.008 & 0.010 & 0.006 \\
Effect scale $\beta_c$ & 1,100 & 600 & 560 \\
Mean weekly spend (€K) & 75.8 & 52.5 & 27.5 \\
ROAS & 4.20 & 2.90 & 1.56 \\
\bottomrule
\end{tabular}
\par\vspace{0.8em}
\begin{minipage}{\columnwidth}
\footnotesize
\textit{Note:} $\lambda_c$ per thousand euros of adstocked spend; $\beta_c$ in thousands of euros per unit of saturated adstock. Parameter values are fictional, chosen to lie in the range of published industry benchmarks.
\end{minipage}
\end{table}

\subsection{Measurement layer}\label{measurement-layer}

The variables an analyst actually feeds an MMM come from business
systems --- campaign calendars, pricing databases, tracking pipelines
--- and inherit their errors. The dataset therefore includes, alongside
the ground truth, observed counterparts:

\begin{itemize}
\tightlist
\item
  \textbf{Observed promotion indicator.} True promotion weeks are
  captured with 70 percent recall; 15 percent of non-promotion weeks are
  false positives, mimicking coordination failures between marketing
  calendars and pricing systems. Overall agreement with the true
  indicator is 82.1 percent.
\item
  \textbf{Observed price level.} The true promotional price is
  contaminated with autocorrelated noise (persistence 0.2, innovation
  standard deviation 0.08, clipped to \([0.5, 1.8]\)), producing
  systematic rather than purely random pricing error; the correlation
  with the true price is 0.89.
\end{itemize}

Media spend is observed without error, matching the practical situation
in which spend comes from billing systems while promotion and price data
come from less reliable sources. Market sentiment \(u_t\) has no
observed counterpart at all.

\subsection{Output data}\label{output-data}

Each row of the dataset records the full state of one week: total sales,
the ground-truth decomposition of Equation (1), the latent baseline
components, true and observed promotional variables, and per-channel
spend and true effects. Columns prefixed \texttt{observed\_}, together
with sales and spend, constitute the practitioner's information set; the
remainder exist for evaluation, and \texttt{ground\_truth.json} records
the true parameters and returns machine-readably so that evaluation code
never hardcodes them. The companion notebook documents the full schema.

\section{The reference instance}\label{the-reference-instance}

\begin{figure}
\centering
\pandocbounded{\includegraphics[keepaspectratio,alt={The reference instance at a glance. (A) Total sales (€M) with the ground-truth decomposition into baseline, promotional uplift, and marketing uplift. (B) Weekly media spend by channel (€K); the TV bursts and the pre-promotion ramps in PLA and Meta are visible.}]{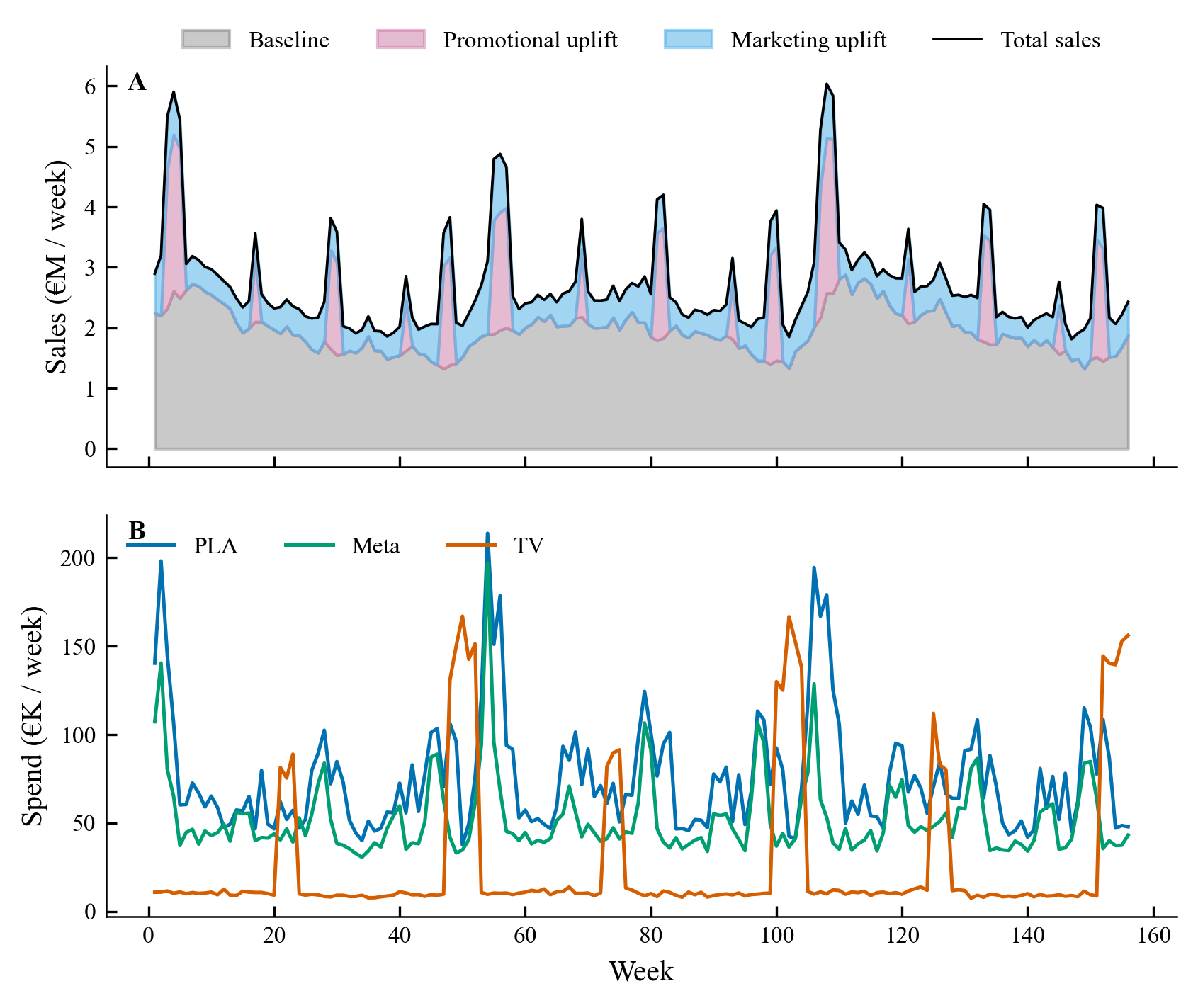}}
\caption{The reference instance at a glance. (A) Total sales (€M) with
the ground-truth decomposition into baseline, promotional uplift, and
marketing uplift. (B) Weekly media spend by channel (€K); the TV bursts
and the pre-promotion ramps in PLA and Meta are
visible.}\label{fig:overview}
\end{figure}

\begin{figure}
\centering
\pandocbounded{\includegraphics[keepaspectratio,alt={The coordination mechanisms, measured in the data. (A) Mean Meta spend by the number of promotional weeks among the next three: spend rises before demand events. (B) PLA spend against the lagged performance ratio the bidding rule observes. (C) The quarterly budget multiplier path.}]{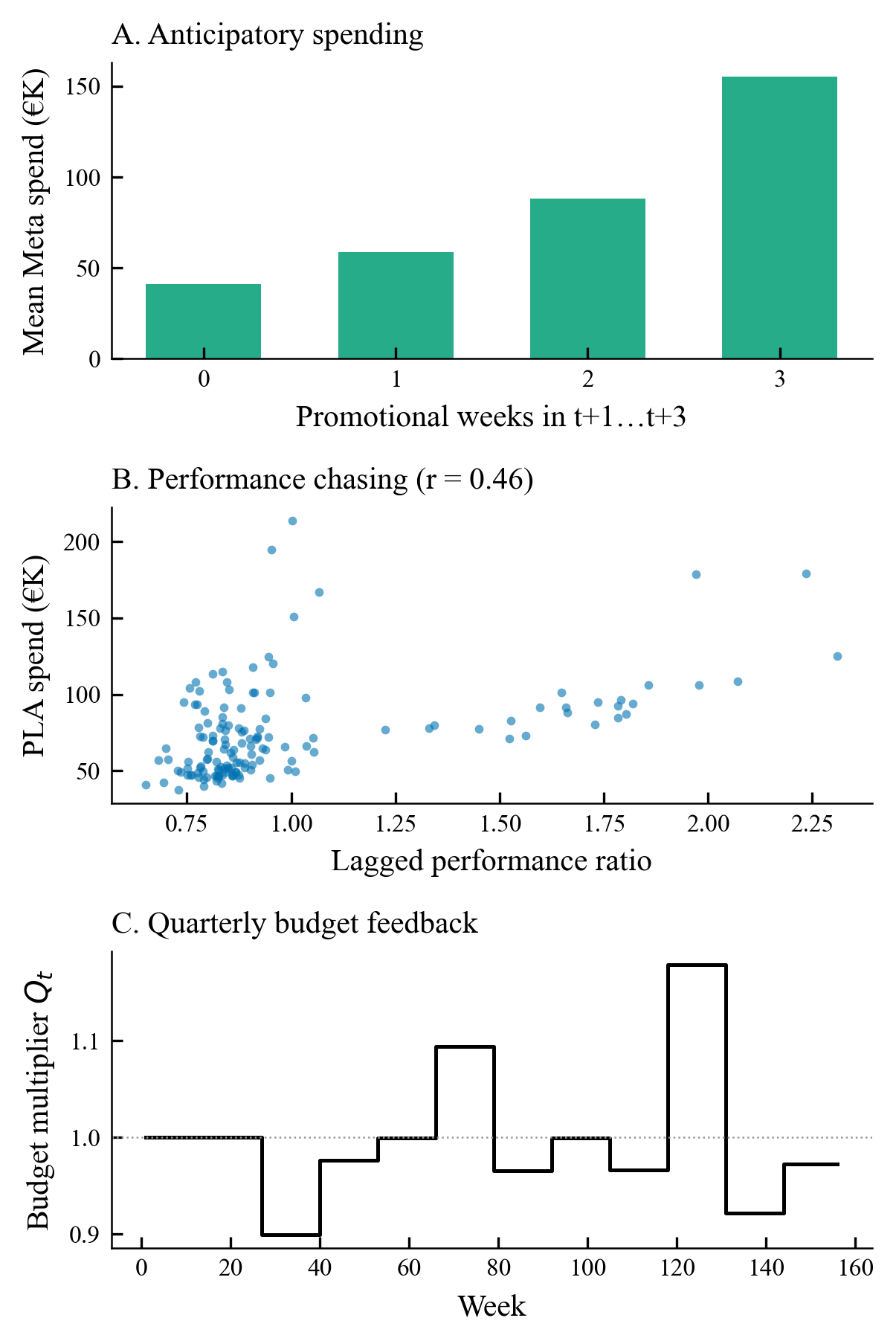}}
\caption{The coordination mechanisms, measured in the data. (A) Mean
Meta spend by the number of promotional weeks among the next three:
spend rises before demand events. (B) PLA spend against the lagged
performance ratio the bidding rule observes. (C) The quarterly budget
multiplier path.}\label{fig:endogeneity}
\end{figure}

The released instance (seed 42) has the following properties. Mean
weekly sales are €2.75 million (minimum €1.81M, maximum €6.03M). Total
media spend over the three years is €24.3 million, an
advertising-to-sales ratio of 5.7 percent. Of cumulative sales, the
baseline accounts for 70.3 percent, promotions for 11.1 percent, and
marketing for 18.7 percent; the blended true ROAS across channels is
3.30. Figure~\ref{fig:overview} shows the sales series with its causal
decomposition and the three spend series; Figure~\ref{fig:endogeneity}
documents the coordination mechanisms empirically, as they appear in the
data.

\section{Geo-experiment extension}\label{geo-experiment-extension}

Models are not the only way firms measure advertising: the market can be
split into comparable region groups, advertising held out of one, and
the effect read off the sales gap between them --- a go-dark
geo-experiment (Vaver and Koehler 2011). Since users of this dataset may
want such experimental data too --- to study experiment analysis, to
study designs (test duration, timing, number of tests) with exact loss
functions, or to combine experimental and observational evidence --- the
generator includes a procedure that simulates them from the same DGP.

The construction is two \emph{parallel universes}. The control universe
is the dataset as planned. In the treatment universe, the tested channel
goes dark during the chosen test windows, and its effect is recomputed
through Equations (7)--(8) over the full history, so the treatment
series embodies the exact causal consequence of going dark --- including
the decay of carryover after each test ends. Opposite-signed noise (one
percent of mean weekly sales) is added to the two universes,
representing week-to-week reallocation of demand between comparable
groups while preserving their sum. (Both universes are kept at country
scale, as if each group covered the whole market, which spares regional
bookkeeping.) The true treatment effect is then known exactly in every
week, during the test and through the post-test carryover, so any
analysis --- of lift, or of the full time path of the treatment--control
gap --- can be scored against exact truth.

\begin{figure}
\centering
\pandocbounded{\includegraphics[keepaspectratio,alt={A simulated go-dark geo-experiment for PLA (test weeks shaded). (A) Sales in both universes (€M). (B) PLA spend (€K): the treatment universe goes dark. (C) The weekly sales difference (€K) against the true effect gap (dashed): zero through the pre-period, opening gradually as adstock depletes, and closing during cooldown.}]{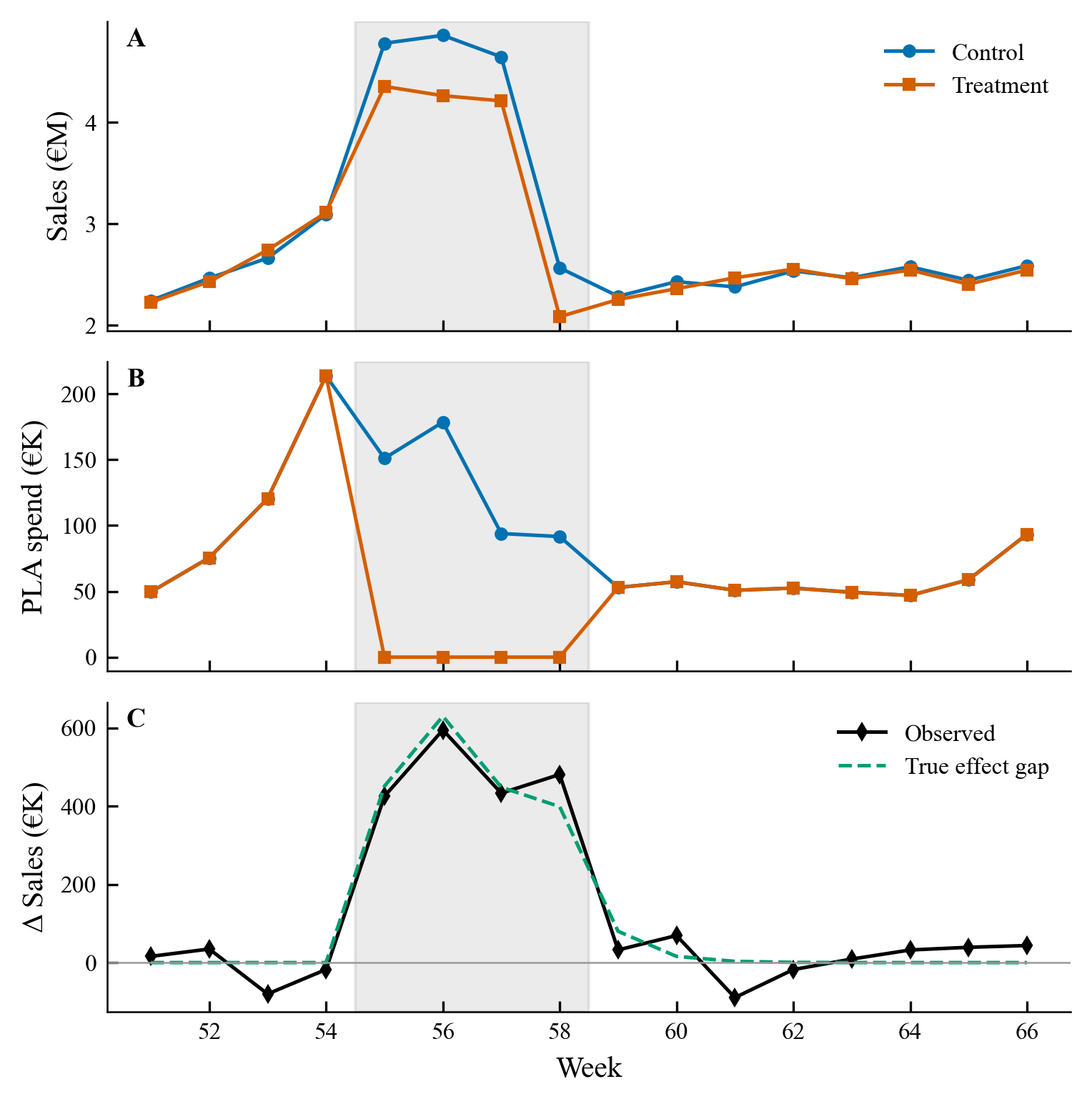}}
\caption{A simulated go-dark geo-experiment for PLA (test weeks shaded).
(A) Sales in both universes (€M). (B) PLA spend (€K): the treatment
universe goes dark. (C) The weekly sales difference (€K) against the
true effect gap (dashed): zero through the pre-period, opening gradually
as adstock depletes, and closing during cooldown.}\label{fig:geo}
\end{figure}

One implementation subtlety is worth recording. The recomputation must
run over the full history, not the analysis window: recomputing on a
windowed series, whose spend history is implicitly zero before the
window, understates the treatment universe's adstock at the window edge
and injects a spurious control--treatment gap into the pre-period. The
generator instead recomputes on the full series and asserts that the
pre-period gap is exactly zero before noise. Figure~\ref{fig:geo} shows
a four-week PLA test: the gap is flat at zero through the pre-period,
opens gradually as the treatment universe's adstock depletes, and closes
during cooldown as carryover fades.

\section{Intended use and
limitations}\label{intended-use-and-limitations}

The dataset supports benchmarking MMM estimators against known effects
under realistic spend coordination; studying methods that combine
observational data with the simulated geo-experiments; and studying
experiment design itself with exact loss functions. Because each
coordination mechanism is a separate function in the generator, ablation
studies --- switching mechanisms off one at a time --- can attribute an
estimator's behavior to specific causes.

Several limitations should be stated. First, the effect structure in
Equation (1) is additive across channels; there are no synergies, and no
competitive or supply-side responses. Second, media parameters are
constant over time, whereas real effectiveness drifts. Third, the PLA
bidding rule responds to pre-media sales, understating algorithmic
feedback (Section 3.4). Fourth, the geo-extension derives both
experimental groups from the national series; it does not model regional
heterogeneity in demand or media response. Future versions could add
channel interactions, time-varying parameters, and genuinely regional
data.

\section{Availability}\label{availability}

The parameterized generator for the seeded reference instance
(\texttt{synthetic\_mmm\_data.csv}) is available in the
\href{https://www.github.com/niklas-heusch/mmm-materials}{mmm-materials
GitHub repository}.

\section*{References}\label{references}
\addcontentsline{toc}{section}{References}

\protect\phantomsection\label{refs}

\end{document}